\documentclass[journal,twocolumn,twoside]{IEEEtran}%
\usepackage{color}

\usepackage{cite}

\ifCLASSINFOpdf
\else
\fi
\usepackage[cmex10]{amsmath}
\usepackage{amssymb}
\usepackage{mathtools}
\usepackage{amsthm}
\usepackage{amsmath}
\usepackage{graphicx}
\usepackage{bm}
\usepackage[belowskip=-2pt,aboveskip=-2pt,small]{caption}

\graphicspath{{figures/}}
\usepackage{epstopdf}

\begin{document}
%
\title{Enabling Sphere Decoding for SCMA} 
%
%
%
\author{Monirosharieh Vameghestahbanati,~
Ebrahim Bedeer,~\IEEEmembership{Member,~IEEE,} Ian Marsland,~\IEEEmembership{Member,~IEEE,} Ramy H. Gohary,~\IEEEmembership{Senior Member,~IEEE,} Halim Yanikomeroglu,~\IEEEmembership{Fellow,~IEEE}\thanks{Manuscript received May 20, 2017; revised August 18, 2017; accepted August 22, 2017. This work is supported in part by an Ontario Trillium Scholarship, in part by Huawei Canada Co., Ltd., and in part by the Ontario Ministry of Economic Development and Innovations Ontario Research Fund - Research Excellence (ORF-RE) program.

 M. Vameghestahbanati, I. Marsland, R. H. Gohary, and H. Yanikomeroglu are with the Department of Systems and Computer Engineering, Carleton University, Ottawa, ON, Canada (email: \{mvamegh, ianm, gohary, halim\}@sce.carleton.ca).
  E. Bedeer is with the School of Engineering, Ulster University, Jordanstown, UK (email: e.bedeer.mohamed@ulster.ac.uk).}}
\IEEEpubid{0000--0000/00\$00.00~\copyright~2017 IEEE}


\maketitle
\pagenumbering{gobble}
\begin{abstract}
In this paper, we propose a reduced-complexity optimal modified sphere decoding (MSD) detection scheme for SCMA.
As SCMA systems are characterized by a number of resource elements (REs) that are less than the number of the supported users, the channel matrix is rank-deficient, and sphere decoding (SD) cannot be directly applied. Inspired by the Tikhonov regularization, we formulate a new full-rank detection problem that it is equivalent to the original rank-deficient detection problem for constellation points with constant modulus and an important subset of non-constant modulus constellations. By exploiting the SCMA structure, 
 the computational complexity of MSD is reduced compared with the conventional SD. We also employ list MSD to facilitate channel coding. Simulation results demonstrate that in uncoded SCMA systems the proposed MSD achieves the performance of the optimal maximum likelihood (ML) detection. Additionally, the proposed MSD benefits from a lower average complexity compared with MPA.
\end{abstract}

\begin{IEEEkeywords}
Sparse code multiple access (SCMA), modified list sphere decoding (MSD), maximum likelihood (ML).
\end{IEEEkeywords}

\IEEEpeerreviewmaketitle

\section{Introduction}
%
%
%
%
\IEEEPARstart{T}{he} need to accommodate diverse types of users and applications necessitates more efficient ways to use the spectrum in 5G systems. 
Sparse code multiple access (SCMA) 
is a non-orthogonal multidimensional codebook-based configuration that can cope with the requirements of 5G systems. In essence, SCMA is a generalization of low density signature (LDS) signaling, whereby a sparse signature matrix is used to reduce the complexity of the detector at the receiver \cite{scma}. 

One of the main challenges in the design of SCMA systems is to overcome the complexity of the receiver that decodes the data generated from all active users. Inspired by the sparsity of the SCMA codewords, 
\cite{scma} uses a near-optimal message passing algorithm (MPA) to detect the SCMA symbols. 
An MPA-based algorithm was proposed in \cite{wei15} that reduces the detection complexity by assigning larger weight factors to those codewords with larger probabilities. However, this improvement results in a degradation in the block error rate at high signal-to-noise ratio (SNR) regimes. 
In \cite{wei2016low}, a hybrid of list sphere decoding (SD) 
and MPA is used to reduce the complexity of SCMA detection. 
However, the SD detection problem is defined per resource elements (RE) and not on the entire block, and thus does not result in an optimal solution. 

In this paper, we investigate the detection problem of SCMA systems, and propose a reduced-complexity optimal modified SD (MSD) detection scheme. Due to the non-orthogonal nature of SCMA systems, the number of REs is less than the number of active users. As such, the channel matrix is rank-deficient, and SD cannot be directly applied.  
To tackle this issue, 
we use Tikhonov regularization \cite{tikhonov} to formulate a new full-rank detection problem that is equivalent to the original rank-deficient detection problem for constellation points with constant modulus, and an important subset of non-constant modulus constellations. 
The complexity of the proposed MSD scheme is reduced by exploiting the sparsity of SCMA codebooks, and the fact that each user spreads the same information bits over a few REs alleviates the need for expanding all tree branches. 
We also use list MSD to provide soft-outputs to be used with channel coding. Simulation results confirm that in uncoded scenarios the proposed MSD scheme achieves the performance of the optimal maximum likelihood (ML) detection. Furthermore, the proposed MSD benefits from a lower average complexity compared with MPA. 
\IEEEpubidadjcol
\section{System Model}\label{sec:sysmodel}
Consider an uplink SCMA system with $K$ users and $N$ orthogonal resource elements (REs), where $N<K$, and each user is connected to $d_v\ll N$ REs only. 
In an $M$-ary signal constellation, each $L_M=\log_2 M$ bits 
  {is mapped to a $d_v$-dimensional complex constellation symbol, $\bm{x}_k=\left(x_{1,k},\dots,x_{d_v,k}\right)^{\textrm{T}}$ that is selected from a $d_v$-dimensional complex codebook $\bm{X}_k$ of size $M$, and defined within a constellation set, $\mathcal{X}_k\subseteq \mathbb{C}^{d_v}$.
  The $N\times d_v$ binary mapping matrix\footnote{The indicator matrix $\bm{P}$ is an $N \times {K}$ matrix with each of its columns defined as $\bm{p}_k =\textrm{diag}\left(\bm{S}_k\bm{S}_k^{\textrm{T}}\right)$. Note that \cite{wei2016low} uses $\bm{p}$ to apply SD on each RE independently, which does achieve the ML performance. Also, using $\bm{P}$ to apply SD on all REs jointly is very challenging and may not be possible. In contrast, $\bm{S}$ allows us to apply SD on all REs jointly, and achieves the ML performance. The increase in dimensionality from $K$ to $K'$ due to using $\bm{S}$ instead of $\bm{P}$ will be compensated in the proposed algorithm.} 
  of user $k$ is denoted by $\bm{S}_k$, where $s_{n,l}=1$, $n\in\left\{1,\dots,N\right\}$ and $l\in\left\{1,\dots,d_v\right\}$, if and only if the $l$th symbol of user $k$ occupies resource $n$. 
 We assume each user consists of $d_v$ layers that is connected to one RE only. The total number of layers is $K'=d_v\:K$, and the mapping matrix of the SCMA code, ${\bm{S}}=\left[\bm{S}_1 \hdots \bm{S}_K\right]$, is then an $N \times {K'}$ matrix with only one non-zero element in each of its columns. 
The set of layers occupying resource $n$ is specified by the position of $1$s in the $n$th row of $\bm{S}$, and is represented by $\mathcal{F}_n=\left\{k'|s_{n,k'}=1\right\}$, $k'\in\left\{1,\dots,K'\right\}$, with cardinality $d_f=\left|\mathcal{F}_n\right|$. 
As we will discuss later in Section III-A and V, to simplify the detector, it is very advantageous for $\bm{S}$ to be an upper-triangular matrix, so the first $N$ columns of $\bm{S}$ constitutes an identity matrix. This can be achieved in scenarios with static resource allocation (RA) by assigning the first $d_v$ REs to the first user, the second $d_v$ REs to the second user, and so on. In scenarios with  dynamic RA, 
it is easy to make $\bm{S}$ an upper-triangular matrix during the RA 
phase by relabelling the REs and users, provided that there exists $N/d_v$ orthogonal users\footnote{On the occasion that there are not $N/d_v$ orthogonal active users in the system, we can relabel the layers rather than the users and some minor modifications to the SD algorithm will be needed.}.

In an uplink transmission scenario over Rayleigh frequency flat fading contaminated by additive white Gaussian noise (AWGN), the $N \times 1$ received signal vector is represented by
   \begin{eqnarray}\label{r}
  \bm{y}&=&\sum\limits_{k=1}^{K}\bm{S}_k \:\bm{H}_k\:\bm{x}_k+\bm{w}\nonumber \\
  &=&\bm{G}\bm{x}+\bm{w},
  \end{eqnarray} 
where $\bm{H}_k = \textrm{diag}\left(h_{1,k},\hdots,h_{d_v,k}\right)$ is a $d_v \times d_v$ diagonal matrix containing the complex channel gains for the $d_v$ REs used by user $k$,
$\bm{x}=\left(\bm{x}_1^{\textrm{T}},\hdots,\bm{x}_K^{\textrm{T}}\right)^{\textrm{T}}=\left({x}_1,\dots,{x}_{K'}\right)^{\textrm{T}}$ is a $K'$-dimensional vector containing all transmitted symbols of all users, and $\bm{w}~\thicksim\mathcal{CN}\left(0,\sigma^2 \mathbf{I}\right)$ is the $N$-dimensional complex Gaussian ambient noise. 
Moreover, $\bm{G}=\left(\bm{g}_1,\dots,\bm{g}_K\right)$, $\bm{g}_k=\bm{S}_k\: \bm{H}_k$, is the $N\times K'$ effective channel gain matrix.

After the reception of $\bm{y}$, a multiuser detection technique is employed to recover each user's codeword $\bm{x}_k$.
The optimal ML detection for SCMA transmitted codewords is given by
\begin{equation}\label{ML}
  \bm{\hat x}=\arg \mathop {\min }\limits_{{\bm{x} \in \mathcal{X}}}\left\lVert
 \bm{y}-\bm{G}\bm{x}
\right\rVert^2,
\end{equation}
 where $\bm{\hat x}=\left({\hat x}_1,\dots,{\hat x}_{K'}\right)^{\textrm{T}}$, denotes the detected symbols, and $\mathcal{X}=\mathcal{X}_1 \times \hdots \times \mathcal{X}_K$, $\mathcal{X}\subseteq \mathbb{C}^{K'}$, contains the constellation set of all users. 
Since the ML implementation is prohibitively complex, we propose a reduced-complexity MSD detection scheme that 
is able to achieve the optimal ML performance. 
\section{MSD Detection Scheme}\label{sec:SD}
In this section, we develop a reduced complexity SD detection scheme that is based on a modified tree search method, and is capable of achieving the ML performance. 
\subsection{Problem Formulation}\label{sec:probformulation}
It is clear that (\ref{ML}) represents an under-determined system; thus, SD cannot be directly applied. To overcome this issue, and inspired by Tikhonov regularization \cite{tikhonov}, we rewrite $\bm{G}$ in \eqref{r} as, $\bm{G}=\left[
\bm{G}^{\left(1\right)}_{N \times N}~\bm{G}^{\left(2\right)}_{N \times \left({K'}-N\right)}\right]$.
We then define the modified effective channel gain matrix as
\begin{eqnarray}\label{g2}
  \tilde{\bm{G}}_{K' \times K'}=\begin{bmatrix*}[l]
{\bm{G}}^{\left(1\right)}_{N \times N} & {\bm{G}}^{\left(2\right)}_{N \times \left(K'-N\right)}\\
 \mathbf{0}_{\left(K'-N\right) \times N}& \mathbf{I}_{\left(K'-N\right) \times \left(K'-N\right)}
\end{bmatrix*}.
\end{eqnarray}
Also, by rewriting $\bm{x}$ in \eqref{r} as, $\bm{x}^{\textrm{T}}=\left[ {\bm{x}}^{\left(1\right)}_{N \times 1}~
 {\bm{x}}^{\left(2\right)}_{\left(K'-N\right) \times 1}\right]^{\textrm{T}}$,
the modified received signal vector can be written as
 \begin{IEEEeqnarray}{rcl}\label{rbar}
\tilde{\bm{y}}&{}={}&\tilde{\bm{G}}{\bm{x}}+\tilde{\bm{w}}  \nonumber \\
 \begin{bmatrix}
 \bm{y}_{N \times 1}\\
 \mathbf{0}_{\left(K'-N\right) \times 1}
 \end{bmatrix} &{}={}& \tilde{\bm{G}} \begin{bmatrix}
 {\bm{x}}^{\left(1\right)}_{N \times 1} \\
 {\bm{x}}^{\left(2\right)}_{\left(K'-N\right) \times 1}
 \end{bmatrix} + \begin{bmatrix}
 \bm{w}_{N \times 1}\\
 - \:{\bm{x}}^{\left(2\right)}_{\left(K'-N\right) \times 1}
 \end{bmatrix}.\IEEEeqnarraynumspace
   \end{IEEEeqnarray}
 As such, the ML detection problem is
 \begin{equation}\label{ML3}
  \bm{\hat{{ x}}}=\arg \mathop {\min }\limits_{{{\bm{x}} \in \mathcal{X}}}\left(\left\lVert\tilde{\bm{y}}-
                                \tilde{\bm{G}}{\bm{x}}\right\rVert^2-\normalsize{\left\lVert{\bm{x}}^{\left(2\right)}\right\rVert}^2\right),
\end{equation}
which represents well-defined systems of equations, and is equivalent to the ML detection problem in (\ref{ML}) for constellation points with constant modulus
, i.e., $\left\|{\bm{x}}^{\left(2\right)}\right\|^2$ is a constant (e.g., 4-QAM or the codebook in \cite{codebook}). Since $\bm{S}$ is an upper-triangular matrix, $\tilde{\bm{G}}$ is upper-triangular and MSD can then be \emph{directly} applied. That is, the choice of $\bm{S}$ alleviate the need of QR factorization prior to SD as in e.g., \cite{hassibi}, which will substantially reduce the computational complexity of the proposed MSD.
\subsection{Modified Tree Search Method}\label{sec:tree}
The proposed MSD detection scheme can be visualized by a search over a tree with $K'$ layers. 
The tree search is performed in  descending order from the last layer  down to the first layer, wherein the layers with the indices $k' \in [\left(k-1\right)d_v+1, \: kd_v]$ correspond to user $k$.
Since each user spreads the \emph{same} information bits over $d_v$ REs, there exists up to $M$ branches for layers with indices $k'=k \: d_v$, $k \in \{1, \hdots, K\}$, and only one branch for layers with indices $k' \in \{\left(k-1\right)d_v+1, \hdots, k\:d_v - 1\}$.~
The modified tree search method is illustrated in Fig. \ref{tree}, for a scenario with $K=6$, $d_v=2$, and $M=4$. 
 \begin{figure}[!t]
\centering
\includegraphics[width = 0.2\textwidth]{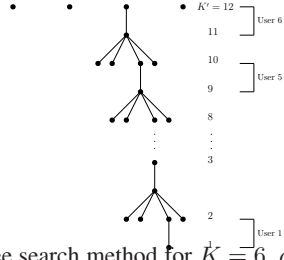} 
 \caption{Modified tree search method for $K=6$, $d_v=2$, and $M=4$.}
\label{tree}
\end{figure}
\vspace{-1 mm}
  \subsection{MSD Extension to Non-constant Modulus Constellations}\label{sec:GMSD}
 {As mentioned in Section \ref{sec:probformulation}, \eqref{ML3} is equivalent to \eqref{ML} for constellation points with a constant modulus, e.g., 4-QAM. However, as long as the SCMA codebook can be written as $\bm{x} = \bm{\Omega}\: \bm{x'}$, where the entries of $\bm{x'}$ are taken from a constant modulus constellation and $\bm{\Omega}$ is a block diagonal matrix, then the decoding algorithm can readily be employed the same way as for the constant modulus case. This condition applies for most good SCMA codebooks, e.g., \cite{taherzadeh14} or 16-QAM. For illustrative purposes, consider an $M$-QAM constellation, $\bm{v}=\left[v_1,\dots,v_M\right]$, with $M = 4^m$, that can be constructed from 4-QAM constellations. Let $\bm{V'}$ represent the $m\:\times\:4^m$ matrix resulting from the Cartesian product of $m$ tuples of 4-QAM constellations, and let $\bm{v'}_i$, $i\in\left\{1,\dots,4^m\right\}$, denote the $i^{\text{th}}$ column of $\bm{V'}$. Each $M$-QAM constellation point, ${v}_i$, is ${v}_i=\sum\nolimits_{j=1}^{m}2^{j-1}\:v'_{ij}$.
~In a similar vein, the $K'$-dimensional vector, $\bm{x}$, containing the transmitted symbols of all layers can be decomposed as $\bm{x} = \bm{\Omega}\: \bm{x'}$, where $\bm{\Omega}=\textrm{diag}\left(\bm{\omega}_1,\dots, \bm{\omega}_{K'}\right)$ is a $K'\times m\:K'$ block diagonal matrix, and $\bm{\omega}_{k'}=\left[2^{m-1},\dots,1\right]$. Further, $\bm{x'}= (\bm{x}_{1}^{'\textrm{T}},\dots,\bm{x}_{K'}^{'\textrm{T}})^{\textrm{T}}$ is an $m\:K'$- dimensional vector, and the entries of $\bm{x'}_{k'}$ are chosen from the 4-QAM constellation. The received signal vector $\bm{y}$ can be re-written as $\bm{y} = \bm{G'}\: \bm{x'}+\bm{w}$,
 where $\bm{G'}=\bm{G}\: \bm{\Omega}$. The MSD detection scheme then runs the same way as the constant modulus case.} 
\section{Complexity Discussion} \label{sec:complexity}
The principles of the SD algorithm necessitate the detected symbols, $\hat{\bm{x}}$, falls within a hypersphere of radius $d$ by ensuring that the following is held \cite{hassibi}:
\begin{IEEEeqnarray}{rcl}\label{ddash}
  d^{2}\:&\geq&\:\sum\nolimits_{i=1}^{K'}\left|{y}_{i}-\sum\nolimits_{j=i}^{K'}\tilde{g}_{i,j}\hat{{x}}_{j}\right|^2, 
\end{IEEEeqnarray}
where $\tilde{g}_{i,j}$ denotes each element of $\tilde{\bm{G}}$.
 From the choice of $\bm{S}$ to be an upper-triangular matrix and from Section \ref{sec:probformulation}, the following are observed for $\tilde{\bm{G}}$, which introduce a reduction in the number of operations involved in detecting the transmitted symbols: Firstly, since the first $N$ columns of $\bm{S}$ form an identity matrix, $\bm{G}^{\left(1\right)}_{N\times N}$ is a full-rank matrix 
 with non-zero diagonal elements, i.e., $ \tilde{g}_{j,j}\neq 0, j=1,\dots,K'$. Secondly, due to the sparsity of  SCMA codebooks, the number of non-zero elements in the first $N$ rows of $\tilde{\bm{G}}$ is $d_f$, and the position of those non-zero elements is determined by the position of ones in the corresponding row of the mapping matrix, ${\bm{S}}$. Thus, only $d_f$ layers are involved in detecting each symbol corresponding to the first $N$ rows of $\tilde{\bm{G}}$. That is, from \eqref{ddash}, to detect the symbol corresponding to the first $N$ layers, i.e., $k' \in \{1, \hdots, N\}$, MSD selects a constellation point, $\hat{x}_{k'}$, that satisfies
\begin{equation}\label{ddashk2}
 d^{2}\:\geq\: d_1^2+\left|{y}_{k'}-\sum\nolimits_{j \in \mathcal{F}_{k'}}\tilde{g}_{k',j}\hat{{x}}_{j}\right|^2,
\end{equation}
where $\mathcal{F}_{k'}=\left\{i|s_{k',i}=1\right\}$.
 Thirdly, due to the presence of $\mathbf{I}$ in \eqref{g2},
 the last $K' - N$ rows of $\tilde{\bm{G}}$ will have only one non-zero element, which is on the diagonal. That is, for the last $K'-N$ layers, i.e., $k'\in \left\{N+1,\dots,K'\right\}$, 
\begin{equation}\label{ddashk3}
d^{2}\:\geq\: d_1^2+\left|{y}_{k'}-\tilde{g}_{k',k'}\hat{{x}}_{k'}\right|^2,
\end{equation}
where $d_1^2=\sum\nolimits_{i\:=\:k'+1}^{K'}\left|{y}_{i}-\sum\nolimits_{j\:=\:i}^{K'}\tilde{g}_{i,j}\hat{{x}}_{j}\right|^2$,
when $k'\in\left\{1,\dots,K'-1\right\}$, and $d_1^2=0$ when $k'=K'$. 
Let $N_{v_1}$ and $N_{v_2}$ denote the average number of visited layers \cite{hassibi,cui13,wang09,mansour14} for $k' \in \{1, \hdots, N\}$, and $k' \in \{N+1, \hdots, K'\}$, respectively. The average complexity of MSD based on \eqref{ddashk2}--\eqref{ddashk3}, and the complexity of $\log$-MPA \cite{scma} using $\mathop {\max }\limits^*$ operation \cite{Hochwald03} is provided in Table \ref{table:comp}.
~Note from \eqref{ddash}, unlike \eqref{ddashk2} and \eqref{ddashk3}, the detection of each symbol involves the contribution of all other symbols. Further, the conventional SD branches up to $M$ possibility for all the $K'$ layers, whereas from Section \ref{sec:tree}, we branch up to $M$ possibilities only for $K$ layers, and branch only 1 possibility for the $K'-K$ layers. This suggests a substantial reduction in the complexity of the proposed MSD compared with the conventional SD{\footnote{As a result, the dimensionality increase due to using $\bm{S}$ instead of $\bm{P}$ is compensated with the MSD scheme.}}.
\begin{table*}[!htp]
\caption{Average complexity of MSD and MPA} 
\centering 
\begin{tabular}{ccc } 
\multicolumn{3}{c}{} \\ [0.5ex]
\hline 
  &MSD& $\log$-MPA  \\ 
 \hline
 Real Summators & $\left(4\:d_f+2\right)\:N_{v_1}+2\:N_{v_2}$&$M\:N\:d_f\left(M^{d_f-1}\left(4\:d_f-2+N_i\left(2+\frac{1}{M}\right)\right)+N_i\left(2-\frac{1}{d_v}\right)+5\right)$ \\
 Real Multipliers & $\left(4\:d_f+2\right)\:N_{v_1}+2\:N_{v_2}$&$M\:N\:d_f\left(4\:d_f\:M^{d_f-1}+5\right)$\\
 $\exp/\log$ Operations & 0&$M\:N\:d_f\:N_i\left(M^{d_f-1}+1\right)+1$\\
 \hline 
\end{tabular}
\label{table:comp} 
\end{table*}
\begin{figure}[!t]
\centering
 \includegraphics[width = 0.33\textwidth]{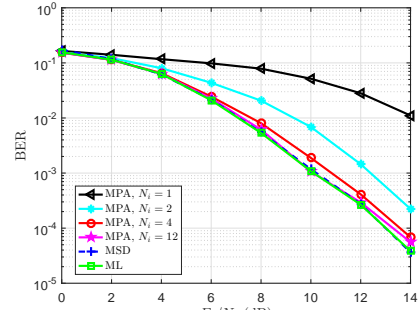}
 \caption{BER performance of 4-ary uncoded SCMA systems over AWGN.}
\label{fig:detection_awgn}
\end{figure}
    \begin{figure*}[t]
\begin{minipage}[b]{0.3\linewidth}
\centering
\includegraphics[width=1.10\textwidth]{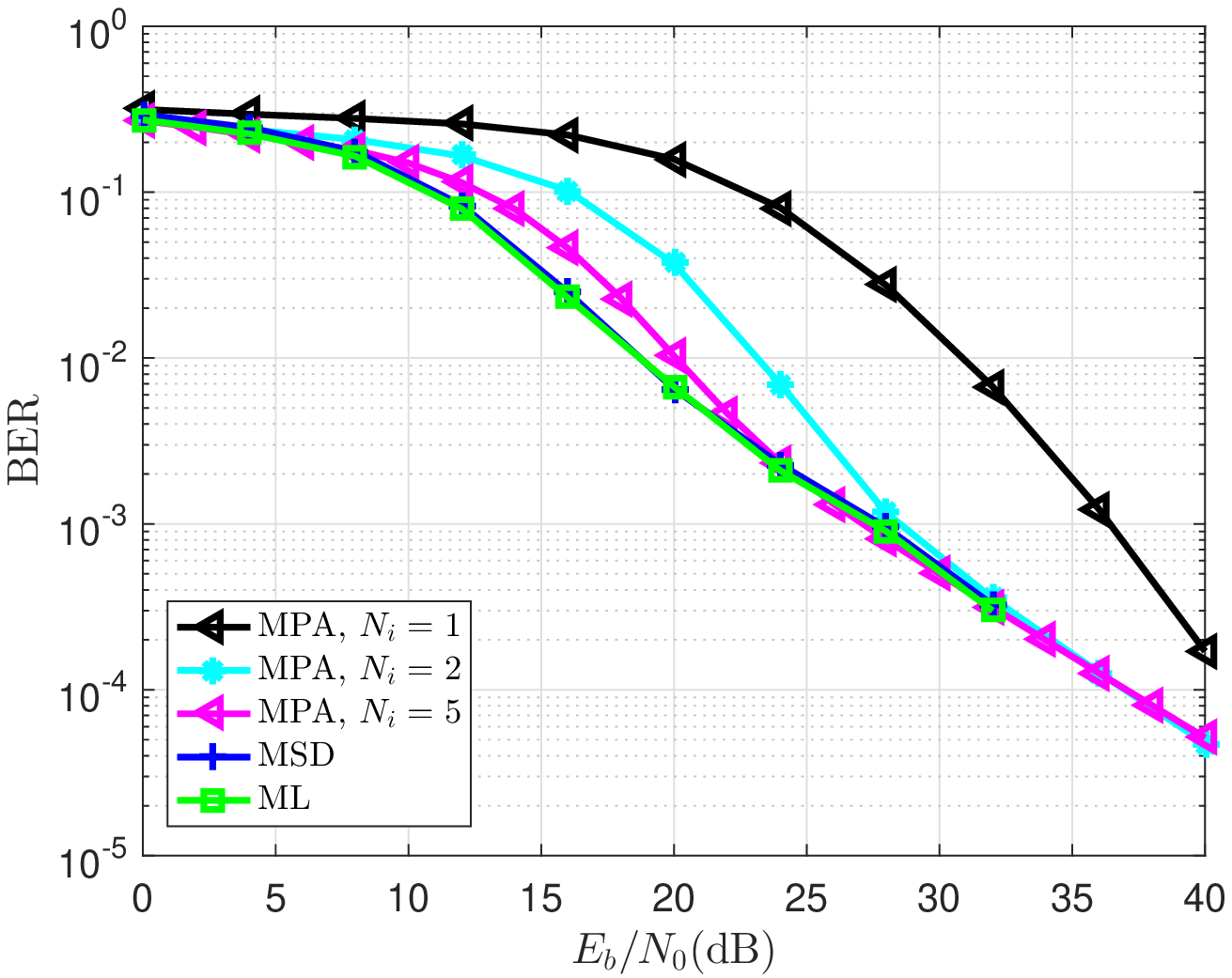}
\caption{BER performance of 16-ary uncoded SCMA systems over fading.}
\label{fig:16ary}
\end{minipage}
\hspace{0.3cm}
\begin{minipage}[b]{0.3\linewidth}
\centering
\includegraphics[width=1.10\textwidth]{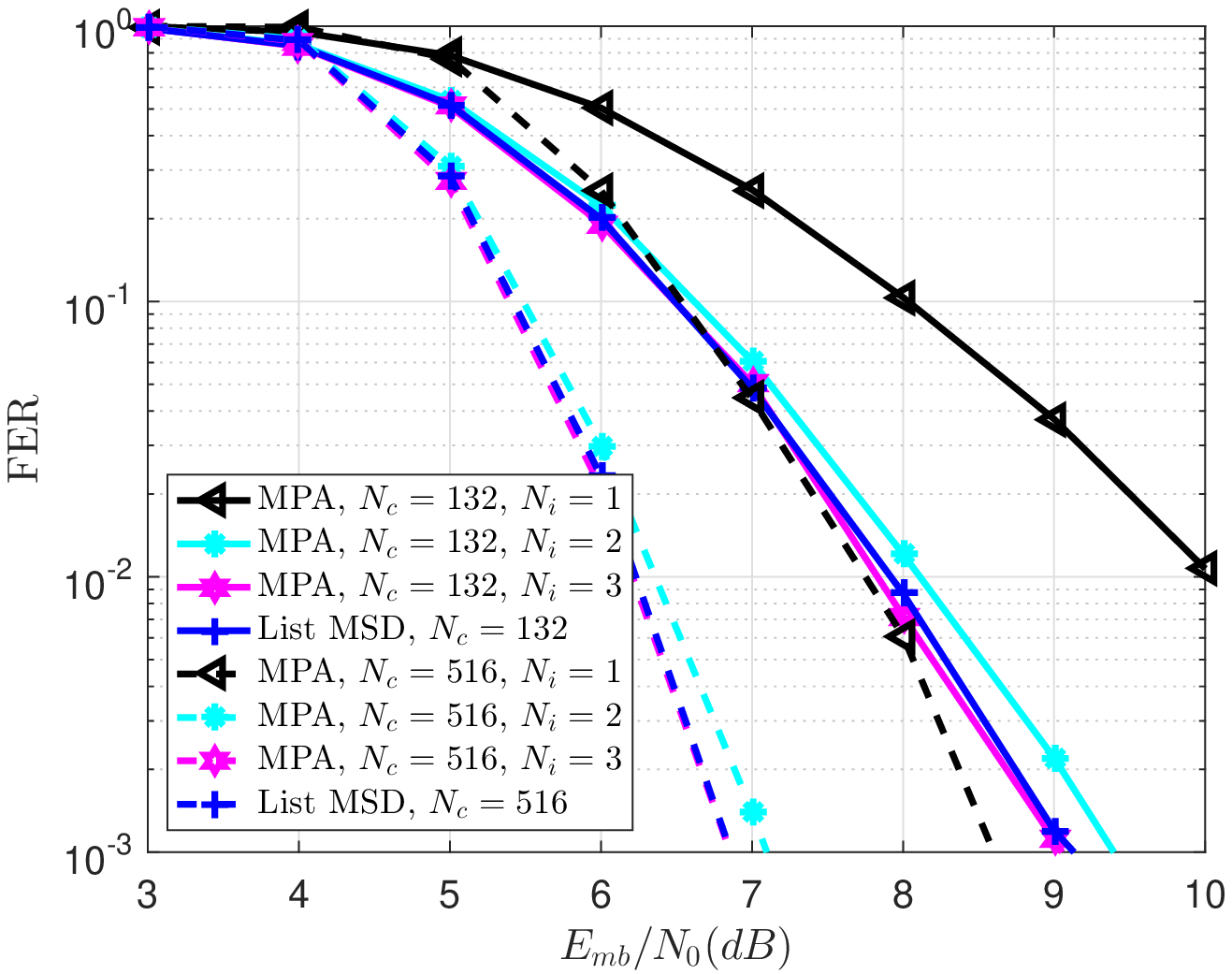}
\caption{FER performance of 4-ary turbo coded SCMA systems over fading.}
\label{fig:coded}
\end{minipage}
\hspace{0.3cm}
\begin{minipage}[b]{0.3\linewidth}
\centering
\includegraphics[width=1.10\textwidth]{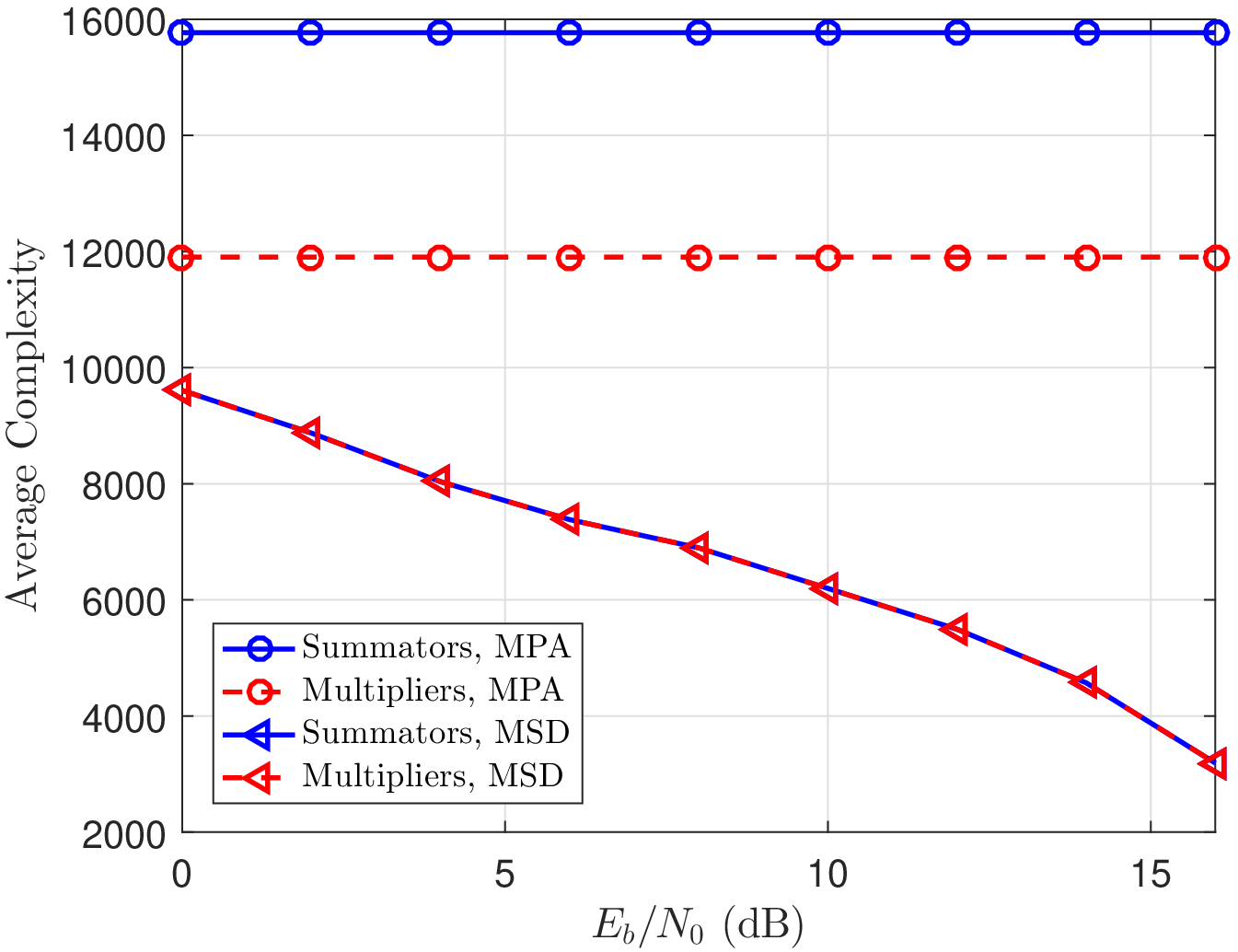}
\caption{The average complexity of an 4-ary uncoded SCMA system over fading.}
\label{fig:flops}
\end{minipage}
\end{figure*}
\section{List MSD} \label{sec:listmsd}
To use channel coding soft outputs are required; we employ list MSD that is based on the list SD \cite{Hochwald03} to provide soft outputs. 
 Let $N_c$ denotes the length of codewords output from the channel encoder. The codewords are partitioned into $N_c/L_M$ digital symbols of $L_M$ bits each. Let $c_k=\left(c_{k,1},\dots, c_{k, L_M}\right)$, $c_{k,m}\in \left(0,1\right)$, be the bits conveyed in one symbol for user $k$, $k\in \left\{1,\dots, K\right\}$. Each $c_k$ is mapped to a $d_v$-dimensional complex constellation symbol, $\bm{x}_k$. That is, $\bm{x}_k=\bm{X}_k\left(c_k\right)$.
To minimize the probability of making a wrong decision on a given bit, its \emph{a posteriori} probability (APP) is maximized.~The log-likelihood-ratio for each bit of each user, $\lambda_{k,m}$, is used to express the APP ratio as

{{\small{
\begin{IEEEeqnarray}{rcl}\label{llrapprox1opt}
  \lambda_{k,m} &{}\approx{}& \mathop {\max }\limits_{\bm{c}\in \mathbb{A}_0}^*\left\{\frac{-1}{2\sigma^2}\left(\left\lVert\tilde{\bm{y}}-
                              \tilde{\bm{G}}{\bm{x}}\right\rVert^2-\normalsize{\left\lVert{\bm{x}}
                              ^{\left(2\right)}\right\rVert}^2\right)+\log\textrm{Pr}\left\{\bm{x}\right\}\right\}\nonumber  \\
                 &&{}-\mathop {\max }\limits_{\bm{c}\in\mathbb{A}_1}^*\left\{\frac{-1}{2\sigma^2}\left(\left\lVert\tilde{\bm{y}}-
                              \tilde{\bm{G}}{\bm{x}}\right\rVert^2-\normalsize{\left\lVert{\bm{x}}
                              ^{\left(2\right)}\right\rVert}^2\right)+\log\textrm{Pr}\left\{\bm{x}\right\}\right\},\nonumber \\
                              \IEEEeqnarraynumspace
  \end{IEEEeqnarray}}}}where $\bm{c}=\left(c_1,\dots,c_K\right)$, $\bm{x}=\bm{X}\left(\bm{c}\right)$, $\bm{X}=\left(\bm{X}_1,\dots,\bm{X}_K\right)$. In addition, $\mathbb{A}_0 = \mathbb{C'}_{k,m},0$ and $\mathbb{A}_1 = \mathbb{C'}_{k,m},1$ denote the set of bits $\bm{c}$ having $c_{k,m}=0$ and $c_{k,m}=1$, respectively, and $\textrm{Pr}\left\{\bm{x}\right\}$ represents the \emph{a priori} probability of $\bm{x}$. The $\mathop {\max }\limits^*$ operation is a numerically stable operation and is defined in \cite{Hochwald03}.
%
  Based on the sign of $\lambda_{k,m}$, we can decide whether that bit corresponds to a binary 0 or 1, i.e., $c_{k,m}=0$ or $c_{k,m}=1$. The magnitude of $\lambda_{k,m}$ also specifies how reliable that specific bit is. Note that computing \eqref{llrapprox1opt} is exponential with the number of users $K$ and the constellation size $M$. That is, for each ${c}_k$, there exists $2^{K\:M}$ hypotheses (all different possibilities of $\bm{c}$ in  $\mathbb{A}_0$ and $\mathbb{A}_1$) to search over in \eqref{llrapprox1opt}. As such, we use list SD to reduce the computational complexity. In the list SD, instead of searching through all hypotheses, we only search over a list $\mathcal{L}$ with $N_{\textrm{cand}}$ most probable hypotheses. That is, in evaluating \eqref{llrapprox1opt},  $\mathbb{A}_0=\bm{c}\in \mathcal{L}\bigcap \mathbb{C'}_{k,m},0$ and $\mathbb{A}_1=\bm{c}\in \mathcal{L}\bigcap \mathbb{C'}_{k,m},1$. 
  More details on list SD are provided in \cite{Hochwald03}.

\section{Simulation Results}\label{sec:sim}\vspace{-2pt}
In this section, we evaluate the performance of the proposed MLSD detection scheme over AWGN and Rayleigh fading channels, where we assume each user observes the same channel coefficients over $d_v$ resources. We provide performance comparisons among different detection schemes for uplink SCMA systems with $K=6$, $N=4$, and $d_v=2$. 

In Fig. \ref{fig:detection_awgn}, we compare the BER performance of uncoded SCMA systems that differ in their multiuser detection techniques, and operate over the AWGN channel, with the 4-ary codebook in \cite{codebook}. We observe that MSD achieves the ML performance. Moreover, MPA approaches ML with increasing number of iterations, $N_i$. In particular, MPA converges to the proposed MSD after 12 iterations in AWGN channels. 
~We depict the performance comparison of uncoded SCMA systems with ML, the proposed MSD, and the MPA detection schemes that operate in Rayleigh fading channels in Fig. \ref{fig:16ary}. All users employ the 16-QAM constellation. As expected, the proposed scheme achieves the same performance as ML, however, the performance of MPA depends on the number of iterations. Particulary, MPA converges to the
proposed MSD after 5 iterations at high SNRs.
Fig. \ref{fig:coded} compares the frame error rate (FER) performance of a rate-1/3 turbo-coded 4-ary SCMA system with the codebook given in \cite{codebook}, and operating over Rayleigh fast fading channels, The list size $N_{\textrm{cand}}$ is set to 600. The turbo codes are chosen from the 3GPP LTE standard \cite{3GPP} with the codeword length of $N_c=132$ and $N_c=516$. As we see, for the two codeword lengths, MPA converges to the list MSD after 3 iterations.
In Fig. \ref{fig:flops}, we show the average number of real summators and multipliers in an uncoded SCMA system with a 4-ary codebook over a  fading channel with MSD and log-MPA. Despite the fact that the complexity of  $\exp/\log$ operation involved in log-MPA is higher than one multiplier, we consider each $\exp/\log$ operation equivalent to one multiplier. As we see in Fig. \ref{fig:flops}, the average complexity of the proposed MSD decreases with the increase in SNR. Although in this scenario MPA converges only after 3 iterations, the proposed MSD benefits from a lower average complexity compared to the widely used MPA.
%

\vspace{-1pt}
\section{Conclusion}\label{sec:conlusion} \vspace{-3pt}
   A reduced-complexity optimal MSD scheme was proposed for SCMA detection that exploits the properties inherited from the structure of SCMA codebooks.
   To facilitate applying the SD algorithm on SCMA systems to achieve the optimal performance, the Tikhonov regularization was used. It was shown that the SCMA detection problem with the modified channel matrix is equivalent to the original rank-deficient detection problem for constellations with constant modulus and an important subset of non-constant modulus constellations.
   The complexity of MSD is reduced compared with the conventional SD based on the sparse structure of SCMA codebooks that does not require to expand all tree branches. In addition, in order to use channel coding, list MSD was employed.
   Simulation results show that in uncoded SCMA systems
   the proposed MSD scheme achieves the performance of the optimal ML detection. Also, the proposed MSD benefits from a lower average complexity compared with MPA. 
\vspace{-2pt}

\bibliographystyle{IEEEtran}
\bibliography{IEEEabrv,mybib}

\end{document}